\documentclass[twocolumn,runningheads,natbib]{svjour2}
\journalname{Astrophysics and Space Science}
\smartqed  
\usepackage{graphicx}
\begin{document} 
\title{Optical spectroscopy of the radio pulsar 
PSR B0656+14 \thanks{Based on observations 
collected at the European Southern
Observatory, Paranal, Chile (ESO Programme 074.D-0512A).}}
\author{S. Zharikov         \and
        R.E.  Mennickent        \and
        Yu. Shibanov         \and
        V. Komarova 
}
\institute{S. Zharikov \at
              OAN IA UNAM, Ensenada, Mexico \\ 
              \email{zhar@astrosen.unam.mx}          
           \and
           R.E. Mennickent \at
              Universidad de Concepcion, Concepcion, Chile
           \and
           Yu. Shibanov \at
           Ioffe Physical Technical Inst. RAS, St. Petersburg, Russia
           \and
           V. Komarova \at
           Special Astrophysical Observatory, RAS, Russia   
}
\date{Received: date / Accepted: date}
\maketitle
\begin{abstract}
We have obtained  the  spectrum of a middle-aged  
PSR B0656+14  in the  4300-9000\AA\ range  
with the ESO/VLT/FORS2. Preliminary results show that 
at 4600-7000\AA\  the spectrum is almost featureless and flat  
with a spectral index   $\alpha_{\nu}$$\simeq$-0.2  that  
undergoes a change  to a positive value at longer wavelengths.
Combining with available multiwavelength data  
suggests two wide, red and blue, flux depressions
whose frequency  ratio is about 2 and    
which could be the 1st and 2nd harmonics
of  electron/positron cyclotron absorption formed 
at magnetic fields $\sim$10$^8$~G  in  upper
magnetosphere of the pulsar. 
\end{abstract}
\section{Introduction}
\label{intro}
Multiwavelength observations of radio pulsars are an important tool 
for the study of not yet clearly understood radiative mechanisms and 
spectral evolution of rotationpowered isolated neutron stars (NSs). 
Optical observations are an essential part of these studies. 
Among a dozen  optically identified   NSs,  only seven pulsars have parallax 
based distances and, therefore,  minimal uncertainties ($\le$10\%)  in 
luminosities.  Their optical spectra are shown in Fig.~1. A reliable
optical spectrum has been obtained  only for the 
young and bright Crab pulsar \citep{Sollerman}.  
 \begin{figure}
\centering
\includegraphics[width=7cm,clip=]{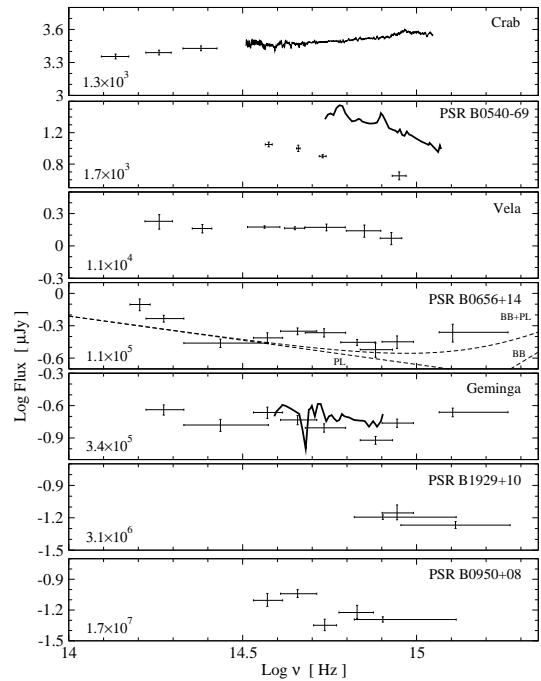}
\caption{Optical spectra of seven pulsars of
different characteristic 
age indicated in the left-bottom corner of each panel in
years. The youngest Crab pulsar is at the top and the oldest
PSR B0950-08 is at the bottom. 
For PSR B0656+14    
dashed lines
show the low energy extensions 
of the blackbody (BB, T=0.84MK) and power law
(PL, $\alpha =0.45$) X-ray spectral components 
and their sum 
\citep{Koptsevich}.
}
\label{fig:1}       
\end{figure}
  Other pulsars are fainter and  
  are represented mainly by broadband 
photometric points. Published  optical spectra  of
 PSR B0540-69 %
\citep{Hill, Serafim} 
are strongly
contaminated  by a bright pulsar  
nebula 
 \citep{Serafim},      
while a tentative spectrum of Geminga  
\citep{Martin} 
is much noisier than available photometric fluxes. \cite{Mignani2}
 reported on the Vela pulsar spectral 
 observations  but these data are not published yet.

In Figure 2 we show   the evolution of luminosity L and radiation
efficiency ${\rm\eta=L/L_{sd}}$ (${\rm L_{sd}}$
is spindown luminosity)   
demonstrated by these seven pulsars 
using the data 
of Table~1 from
\cite{Zhar2}.
We note significantly non-monotonic dependencies
of ${\rm\eta_{Opt}}$ 
and ${\rm\eta_{X}}$  
 versus
pulsar age with a pronounced  minimum at the beginning of the middle-age
epoch (${\rm \simeq10^4}$~yr) and  comparably higher  efficiencies of  younger  
and older pulsars. 

\begin{figure}[t]
\centering
\includegraphics[width=6.5cm,bb=0 0 560 955,clip=]{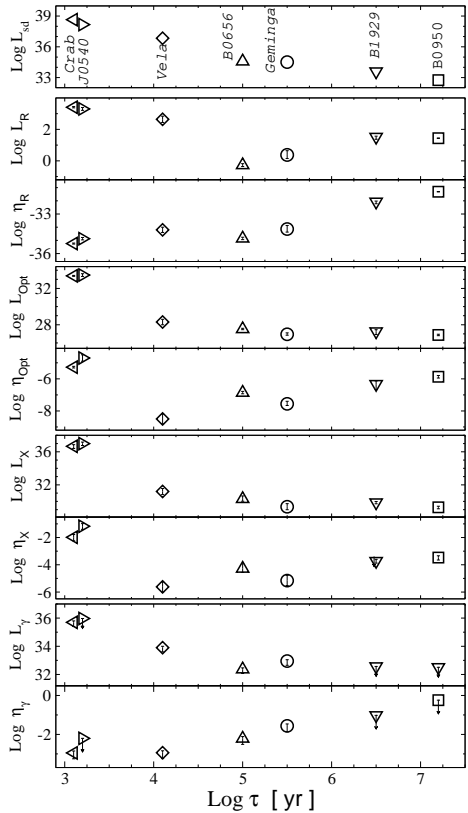}
\caption{{\sl From top to bottom:} 
Evolution  of the spindown, radio, optical, X-ray, and $\gamma$-ray luminosities and respective 
efficiencies  
demonstrated  by the 7 optical pulsars from
Fig.~1.} 
\label{fig:1}       
\end{figure}
Owing to its relative brightness and proximity, the middle-aged PSR
B0656+14 is one of isolated NSs most intensively studied in
different wavelengths.
It was discovered in radio 
by \cite{Manchester}, then 
identified in X-rays with Einstein, and observed in details with 
ROSAT,  ASCA, Chandra and XMM (for references see  
\cite{Shib1, Shib2}). 
The X-ray emission 
can be described as 
a combination of thermal radiation from the entire surface of a cooling
NS and  from hotter 
polar caps
heated by relativistic particles  of magnetospheric
origin. 
An excess over the hot thermal component at energies
$\ge$2 keV was interpreted as nonthermal radiation from the pulsar
magnetosphere \citep{Greiv}. 
 The pulsar has  been also
marginally detected in $\gamma$-rays  ($\ge$50  MeV) 
by 
\cite{Raman}.

 In the optical  PSR B0656+14 was identified  
by 
\cite{caraveo} 
with the ESO/NTT telescopes in the V band. 
 It was then studied in UV with the HST/FOC
 in the F130LP, 
  F430W, F342W and F195W bands \citep{Pavlov1}  
and  with the  HST/WFPC 
in  the  F555W band 
\citep{Mignani1}.  
 Detailed  photometric   studies 
 in the optical-NIR 
  were performed  
 by  
\cite{Kurt}, \cite{Koptsevich}, \cite{Komarova},
and \cite{Shib2}. 
  The studies 
  showed that the bulk of the optical 
   radiation 
   is of  nonthermal origin.
This was confirmed by the detection of  coherent optical
 pulsations with the radio pulsar period  
 in the
 B band
\citep{Shearer},
 in a wide 400-600 nm passband 
\citep{Kern} 
 and in NUV  
\citep{Shib1}. The  
pulse profiles are  rather
sharp 
with a high pulse fraction 
as expected  for  nonthermal emission mechanisms.  
\begin{figure}
\centering
\includegraphics[width=8cm,bb=0 0 760 403,clip=]{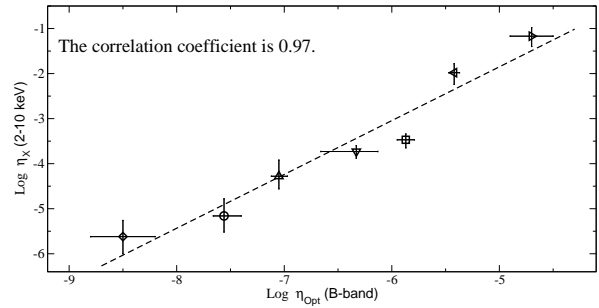}
\caption{Relationship between the B-band optical
and 2-10 keV 
X-ray efficiencies  
for the same 7 pulsars as in Fig.~1,2.}
\label{fig:1}       
\end{figure}
The
phase integrated multiwavelength spectrum
(Fig.~4)   shows  
\citep{Koptsevich,
Shib2}  
 that the NIR-optical-UV spectral energy distribution 
 is,  in a first approach, compatible with the low
energy extension of the sum  of the X-ray thermal blackbody
(BB) spectral 
component from the whole NS surface  and the power law
(PL) component  dominating in the high energy tail. 
The  BB extension does not contribute  at longer  
wavelengths where  the optical-NIR
fluxes are in a good agreement with the PL  alone (Fig.~1). 
This indicates  a common origin of the nonthermal 
optical and X-ray emission, which is strongly supported by 
a good coincidence in phase and
shape of 
the pulse profiles in the
optical and  the X-ray tail 
\citep{Shib1}.  
The  same origin of the nonthermal optical and
X-ray photons is likely to be  a general property for
other pulsars detected in both ranges, as follows from
a strong correlation  between respective  
efficiencies (Fig.~3) found by
\cite{Zhar1,Zhar2}. 

In the optical emission of PSR
B0656+14 there is  an apparent, $\simeq$(3-5)$\sigma$,  
 flux excess  over the PL  
 ``continuum" at Log$(\nu)\simeq$14.7 
 (Fig.~1). This could reveal an additional, 3rd, spectral
 component to the  BB+PL discussed above.  
Here we present first results of the optical
spectroscopy of the pulsar partially motivated
by more detailed studies of the excess. In a broader sense, these results also allow us to
consider, for the first time, 
optical 
properties
of a middle-aged 
ordinary pulsar 
at the spectroscopic level, as it has
been achieved so far only for the Crab pulsar.

\begin{figure*}
\centering
\includegraphics[width=15.cm,clip=]{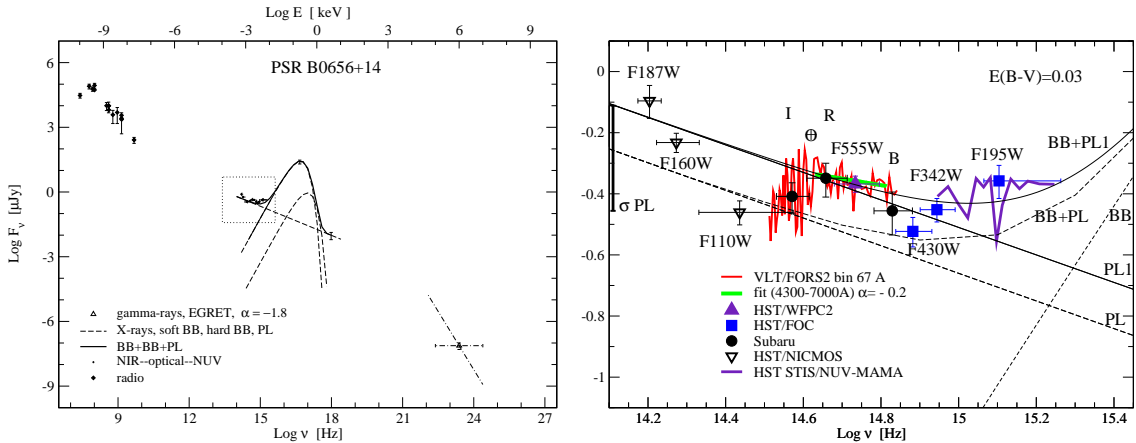}
\caption{{\sl Left:} Unabsorbed multiwavelength spectrum of  PSR B0656+14   
from the radio through $\gamma$-rays.
The box marks the 
range zoomed 
in the right panel.
{\sl Right:} The spectral and photometric fluxes
of PSR B0656+14 in the NIR-optical-NUV range  obtained
with different telescopes and instruments, as notified
in the plot.  
Black dashed lines show low energy extensions   
of the 
soft blackbody (BB) and  power law (PL) spectral fits       
and their     
sum (BB+PL) obtained in X-rays. The black solid lines 
show the same but with the PL normalization shifted up
by a factor of 1.4 (PL1) to fit the upper edge of its
1$\sigma$ error bar shown at the
left side of the plot. These are possibly a better match
for the optical and NUV spectra than the dashed lines.  
The symbol $\oplus$ marks the Earth atmospheric absorption band  near 7600\AA.
}
\label{fig:1}       
\end{figure*}

\section{Observations  and 
data analysis 
}
\label{sec:1}
The spectrum of  PSR B0656+14 was obtained 
on November--December 2004 and
February 2005  
during several observational runs of the ESO
program 074.D-0512A using  the VLT/UT1 
telescope in a service mode.  The
FORS2\footnote{www.eso.org/instruments/fors} 
instrument was used in a long
slit spectroscopic  setup 
with the grating GRIS\_300V and the filter
GG435, which cover the wavelength interval of about
4300-9600\AA\  and  
 provide a medium spectral resolution of  3.35\AA/pixel.   
 The slit width 
was 1$''$ and its position angle was selected in a such way as to
obtain also spectra of several nearby stars for
a sure  pulsar  astrometric 
referencing  and wavelength/flux calibration. 
Eighteen 1400~s  science spectroscopic exposures were taken
with a total exposure time of 25200~s
at a mean seeing  of 0.6$''$.  
Standard  reference frames (biases, darks,  
flatfields, lamps) were obtained  
in each  observational run, while 
the slit and slitless observations  
of  spectrophotometric standards (Feige110, LTT3218 and
LTT1788)  for the flux calibration were carried out in separate  
runs on the same nights. 

 A combination of the MIDAS
and IRAF packages was used for standard CCD data reduction, cosmic-ray
track removing, 
spectra extraction,  and  
subsequent data analysis.
A faint, R=24.65,  pulsar  is at
a limit of spectroscopic capability of the VLT.  
Nevertheless,  excellent seeing conditions 
allowed us to resolve its spectrum  even at  each
individual  exposure,  albeit with a  low signal
to noise ratio S/N.    
These  exposures  were co-added.  The
spectrum  was then extracted with a 3 pixel wide extraction slit 
(0.2$"$/pix) centered on the pulsar. The backgrounds were extracted
with a 6 pixel wide slit centered  above and below the center of 
the pulsar spectrum. 
 The correction factor for the PSF 
and  sensitivity function
were obtained from the Feige110 standard  observations.
The  S/N of the resulting 
spectrum  
was about 4 (per pixel) in the 4450-5500\AA\ range 
and declined to $\sim 1$  near/above
8000\AA,  due to  higher
sky backgrounds  and a  drop in sensitivity 
towards longer wavelengths. 
We binned the spectral flux in 20
pixel bins  (67\AA) to get S/N  near/above 15 and 4, 
respectively,  making the flux accuracy to
be comparable with  that of available
photometric data.

\section{Results and discussion}
\label{sec:3}
The binned and dereddened with ${\rm A_V}$=0.093
spectrum of the pulsar is shown 
in Figure~4 
(red curve in the right panel).     
 We  show also   
available multiwavelength  data (see \cite{Shib2}
for the data and  ${\rm A_V}$ 
description).  
The spectrum
is  in a good agreement  with the 
 broadband VRI 
fluxes, while it is somewhat higher than the B band
flux.  
 It does not show 
any strong nebular emission lines that could be
responsible for the apparent VR  excess,  
%
mentioned  above as a 3rd component, while the presence
of  weak 
features   
can not be completely ruled out. 
The continuum of the dereddened spectrum in 4600-7000\AA\ range
 has a  power law shape $\propto\nu^{-0.2\pm0.2}$
 (green line) 
in agreement  with the nonthermal nature of the bulk of the optical emission. 
Its slope is close to the value expected in this range
from   the BB+PL   extended from
X-rays. 
The slope likely undergoes a change to
a positive value between the R and I bands, as follows also from the
photometric data. 

 Within uncertainties   the bluest end of the optical
 and  the reddest end of the NUV \citep{Shib1} spectra
  are
 compatible with each other, 
 suggesting a smooth connection of  both.
However,  
 the F430W photometric flux in the gap between them 
drops below this connection  
with a significance 
of about 2$\sigma$  of the flux
confidence level. 
Unless this is a
result 
of some unknown  systematics in calibration, 
it suggests a spectral dip in the pulsar emission 
centered near  $\simeq$4300\AA\ ($\simeq$14.83 in Log($\nu$)).

The  optical and NUV spectra and  two NIR  photometric
points, F160W and F187W, almost perfectly match the
 BB+PL  extension 
from X-rays, if 
the PL normalization is taken to be a factor of 1.4 higher (solid lines) 
than its best X-ray fit value (dashed lines). The change  
is  within 1$\sigma$ uncertainty of the  fit.  
Considering the solid line version 
as a new optical ``continuum level" we find an
additional  and   more significant  flux depression in the red
part of the  spectrum    
overlapping  the I  and  F110W  bands 
and 
centered   
near  9000-10000\AA\ (Log($\nu)$$\simeq$14.5).  

Additional spectral studies 
are necessary to confirm the suggested ``red" and ``blue"
features  and  to measure   their
shapes and wavelengths more accurately. 
Nevertheless, if they are real,       
an  approximate   blue-to-red
frequency ratio is $\simeq$2. This indicates that    
 they can be  the  1st and  2nd  harmonics 
of an electron/positron cyclotron absorption formed in
 the upper magnetosphere 
of the pulsar at  an effective altitude 
where the magnetic field B$\simeq$10$^{8}$~G. 
This is $\simeq$360 km, assuming a dipole NS field with  
a surface value of   
4.66$\times$10$^{12}$ G, as derived from spindown
measurements. 
The absorbing $e^{\pm}$  
have to be cooled enough  and provide a sufficient   
optical depth above a source of the nonthermal
continuum.  
The source altitude  is, therefore,  $<$360 km and    
much below  the light cylinder radius of 
$\simeq$18$\times$10$^{3}$ km, likely suggesting its polar cap origin. 
The  features are  broad, as
expected  from 
the magnetospheric field
inhomogeneity.
%
Tentative  
($\leq$2$\sigma$)  absorption features  in the NUV
spectrum  at Log($\nu)\simeq$15 and  15.1  \citep{Shib1}
may be 
the 3rd and 4th 
harmonics,
respectively,  
which are fainter   
as the cyclotron harmonic 
intensity  decreases with its number
${n}$.   

Similar, albeit less significant, spectral features  are likely seen
in  the photometric and spectral data
of another middle-aged pulsar Geminga (Fig.1). 
The absence of strong
nebular 
lines  
suggests  that the features are also of the NS magnetospheric origin.  
 To explain the Geminga spectrum
 \cite{Martin} applied  
a toy model of an ion cyclotron absorption at 
B$\simeq$10$^{11}$~G  in the
inner magnetosphere of the NS combined with
the BB and PL  components. At this  field    
a low ${n}$ ion cyclotron frequencies  indeed 
fall in the optical range.
However, it is likely that the nonthermal optical 
emission is generated  in the upper 
magnetosphere  where the magnetic field is 
by orders of magnitude weaker  and 
any  ion cyclotron absorption of the 
respective optical continuum is negligible. 
In this case the electron/positron cyclotron absorption 
or scattering  
appears to be more plausible
interpretation.   
%
%

The optical spectrum of the young Crab-pulsar is featureless
and has a different  (positive) slope (Fig.~1). The ten
times older Vela-pulsar has a flat and also featureless
spectrum  (Mignani, private communication).
The spectroscopy of  PSR B0656+14  
demonstrates that  optical spectra 
of middle-aged pulsars can be  
distinct from those of  younger ones by the presence of unusual
spectral features  or slope changes. 
To study  this  new spectral observations  of  PSR B0656+14 in the NIR and    
3000-5000\AA\ ranges are needed.
A question, whether the observed  difference in pulsar spectra
is simply caused by different pulsar  geometry 
or by a change of  physical conditions in the
emission region  with  age,
demands   quantitative modeling  physical
processes in pulsar magnetospheres.

\begin{acknowledgements}
The work was partially supported by DGAPA/PAPIIT project IN101506, CONACYT 48493,
RFBR (grants 05-02-16245, 05-02-22003) 
and Nsh 9879.2006.2. REM was supported 
by Fondecyt 1030707. 

\end{acknowledgements}



\end{document}